\documentclass[onecolumn,draft,prd,nofootinbib]{revtex4}
\usepackage{epsf}
\usepackage{dcolumn}
\usepackage{bm}

\newcommand{\be}{\begin{equation}}
\newcommand{\ee}{\end{equation}}

\newcommand{\epm}{\ensuremath{e^{\pm}\;}}

\def\etal{{\it et al.}~}
\def\4he{$^4$He}
\def\3he{$^3$He}
\def\7li{$^7$Li}
\def\Yp{Y$_{\rm P}$~}

\def\mb{$m_{\rm B}$}
\def\nb{$n_{\rm B}$}
\def\rhob{$\rho_{\rm B}$}
\newcommand\la{\lower0.6ex\vbox{\hbox{\ensuremath{\buildrel{\textstyle<}\over{\sim}\ }}}}
\newcommand\ga{\lower0.6ex\vbox{\hbox{\ensuremath{\buildrel{\textstyle>}\over{\sim}\ }}}}

\newcommand{\obh}{\ensuremath{\Omega_{\rm B} h^2\;}}

\begin{document}


\title{The Cosmological Evolution of the Average Mass Per Baryon}

\author{Gary Steigman\\}

\affiliation{Departments of Physics and Astronomy, The Ohio State University,
Columbus, OH 43210\\}

\date{{\today}\\}

\begin{abstract}

Subsequent to the early Universe quark-hadron transition the
universal baryon number is carried by nucleons: neutrons and protons.
The total number of nucleons is preserved as the Universe expands, but 
as it cools lighter protons are favored over heavier neutrons reducing 
the average mass per baryon.  During primordial nucleosynthesis free 
nucleons are transformed into bound nuclides, primarily helium, and 
the nuclear binding energies are radiated away, further reducing the 
average mass per baryon.  In particular, the reduction in the average 
mass per baryon resulting from Big Bang Nucleosynthesis (BBN) modifies 
the numerical factor relating the baryon (nucleon) mass and number 
densities.  Here the average mass per baryon, \mb, is tracked from 
the early Universe to the present.  The result is used to relate the 
present ratio of baryons to photons (by number) to the present baryon 
mass density at a level of accuracy commensurate with that of recent 
cosmological data, as well as to estimate the energy released during 
post-BBN stellar nucleosynthesis. 

\end{abstract}


\maketitle

\section{~Introduction}
\label{intro}

Big Bang Nucleosynthesis (BBN) is regulated by the competition among 
various two-body collisions leading to the creation of complex nuclides 
(mainly helium-4, with trace amounts of deuterium, helium-3, and 
lithium-7) built from neutrons and protons.  The nucleon {\bf number} 
density (\nb) is key to the BBN-predicted primordial abundances.  In 
contrast, for studying the growth and evolution of structure in the 
Universe, the baryon (nucleon) {\bf mass} density (\rhob), including 
the small contribution from the accompanying electrons, plays a more 
direct role.  Of course, the two parameters are directly related 
through the average mass per baryon: \rhob/\nb~= \mb.  Given the high 
precision of recent cosmological data, it is important to match the 
adopted value of the mass per baryon to the accuracy demanded by the 
data when relating the baryon-to-photon ratio to the baryon mass density 
parameter.  It is shown here that as the Universe expands and cools, 
\mb~evolves from a value of \mb~= 939.17 MeV at very early times, shortly 
after the completion of the quark-hadron transition, to a value of 
\mb~= 938.88 MeV just prior to the onset of BBN at $T \approx 70$~keV, 
a few minutes after the expansion began.  The post-BBN average mass 
per baryon has decreased to \mb~= 937.12 MeV, leading to $\eta_{10} \equiv 
10^{10}(n_{\rm B}/n_{\gamma})_{0} = (273.9 \pm 0.3)~\Omega_{\rm B}h^{2}$, 
where $\Omega_{\rm B}$ is the present ratio of the baryon mass density to 
the critical density and $h$ is the present value of the Hubble parameter 
in units of 100 kms$^{-1}$Mpc$^{-1}$.

\section{~The Average Mass Per Baryon}
\label{rhob}

\subsection{Pre-BBN}

Prior to Big Bang Nucleosynthesis, at temperatures below the quark-hadron 
transition, the baryons present in the Universe are nucleons: neutrons 
and protons ($m_{n} = 939.565$~MeV, $m_{p} = 938.272$~MeV~\cite{pdg,audi}).   
In weak equilibrium at high temperatures there are nearly as many neutrons 
as protons: $n/p =$ exp$(-\Delta m/T) \rightarrow 1$ for $T \gg \Delta m 
= 1.2933$~MeV~\cite{pdg}.  In this unachievable limit, \mb~$\rightarrow 
(m_{n} + m_{\rm H})/2 = 939.17$~MeV.  Note that the {\bf hydrogen} mass 
($m_{\rm H} = 938.783$~MeV~\cite{audi}) is used here since in the later 
evolution of the Universe the proton and electron go hand in hand.  
However, even at $T = 100$~MeV, the ratios of baryons to protons ($\equiv$ 
hydrogen $\equiv$~H) by number and by mass are
\be
T = 100~{\rm MeV}:~~n_{\rm B}/n_{\rm H} = 1 + n/p = 1.98715, 
~~~\rho_{\rm B}/\rho_{\rm H} = 1 + 1.000833(n/p) = 1.98797,
\ee
so that the average mass per baryon remains at \mb~= 939.17~MeV.  As 
the Universe cools further, the ratio of neutrons to protons shifts 
even more in favor of the lighter proton, reducing the average mass 
per baryon.  At temperatures below $\sim 1$~MeV the neutron to proton 
ratio deviates from (exceeds) its equilibrium values, and at the onset 
of BBN at a temperature of $\sim$~70 keV, detailed calculations~\cite{ks} 
reveal that $n/p \approx 1/7$.  At this time,  
\be
T = 70~{\rm keV}:~~n_{\rm B}/n_{\rm H} = 1 + n/p \approx 8/7 = 1.14286, 
~~~\rho_{\rm B}/\rho_{\rm H} = 1 + 1.000833(n/p) \approx 1.14298,
\ee
so that the average mass per baryon has now decreased to \mb~$\approx$ 
938.88~MeV.

\subsection{Post-BBN}
\label{postbbn}

During BBN the free neutrons and protons are rapidly transformed 
into light nuclides, resulting in a mixture dominated by hydrogen 
and helium-4, along with trace amounts of deuterium and helium-3 
(any $^3$H produced during BBN decays to \3he).  The mass-7 nuclides 
(\7li, $^{7}$Be) are the only others produced in astrophysically 
interesting abundances, but the BBN-predicted abundance of mass-7 
is so small as to ensure its contribution to the analysis here is 
negligible (for a review of BBN and further references, see Steigman 
(2005)~\cite{s05}).  Since the binding energy of these light nuclides 
has been radiated away, the average mass per baryon has decreased.  
Until stellar processing begins, much later in the evolution of the 
Universe, the baryon mass and number densities are dominated by H, 
D, \3he, and \4he, and the average mass per baryon is frozen at its 
post-BBN value.  

Introducing the abundance ratios by number with respect to hydrogen, 
$y_{i} \equiv n_{i}/n_{\rm H}$, the post-BBN baryon number and mass 
densities, may be related to the corresponding number and mass 
densities of hydrogen,
\be
n_{\rm B}/n_{\rm H} = 1 + 2y_{2} + 3y_{3} + 4y_{4} + ...~,
\ee
\be
\rho_{\rm B}/\rho_{\rm H} = 1 + (m_{2}/m_{\rm H})y_{2} + (m_{3}/m_{\rm H})y_{3} 
+ (m_{4}/m_{\rm H})y_{4} + ...~.
\ee
Since any electrons which remain after \epm annihilation are coupled 
electromagnetically to the charged nuclei, the appropriate masses 
to be used in eq.~4 are the {\it atomic} masses of the neutral atoms.  
The neutral atom masses adopted here from Audi \etal (2003)~\cite{audi} 
are: $m_{2} = 1876.12$~MeV, $m_{3} = 2809.41$~MeV, $m_{4} = 3728.40$~MeV.

As a zeroth approximation, ignore the contributions from D, \3he (and 
\7li), and assume that the primordial \4he abundance is $y_{4} = 1/12$ 
(so that the primordial \4he mass fraction, {\bf defined as} Y$_{\rm P} 
\equiv 4y_{4}/(1+4y_{4})$, is \Yp = 0.250).  This assumes that all 
neutrons available at BBN are incorporated into \4he and corresponds 
to a neutron to proton ratio, evaluated just prior to the onset of BBN, 
$n/p = 1/7$.  In this approximation, $n_{\rm B}/n_{\rm H} = 4/3$ and 
$\rho_{\rm B}/\rho_{\rm H} = 1.3310$, so that \mb~= 937.11 MeV.  For 
arbitrary Y$_{\rm P}$ (see Cyburt (2004)~\cite{cyb} and Serpico \etal
(2004)~\cite{serpico}),  
\be
m_{\rm B}/m_{\rm H} = 1 - (1 - {1 \over 4}({m_{\rm He} \over m_{\rm H}}))
{\rm Y_{P}} = 1 - 0.007119{\rm Y_{P}},
\ee
or
\be
m_{\rm B} = 938.112 - 6.683({\rm Y_{P}} - 0.250)~{\rm MeV}.
\ee
For \Yp = 0.250, this yields \mb~= 937.11 MeV, in perfect agreement with
the result above.

It is possible to improve upon this approximation by utilizing the
BBN-predicted, primordial abundances of the light nuclides.  For example, 
the WMAP-Only, 3-year data from  Spergel \etal \cite{spergel} correspond 
(see \S\ref{eta} below) to a baryon-to-photon ratio of $\eta_{10} = 
6.12^{+0.20}_{-0.25}$.  Using this estimate in the fitting formulae 
for the BBN-predicted abundances from Kneller \& Steigman~\cite{ks}, 
the primordial abundances of the light nuclides may be calculated.  
For $\eta_{10} = 6.12$, the BBN-predicted abundances are: $y_{2} = 
2.56 \times 10^{-5}$, $y_{3} = 1.05 \times 10^{-5}$, and Y$_{\rm P} \equiv 
4y_{4}/(1+4y_{4}) = 0.2482$.  With these abundances, $n_{\rm B}/n_{\rm H} 
= 1.3302$ (in contrast to 4/3 in the zeroth approximation above) 
and $\rho_{\rm B}/\rho_{\rm H} = 1.3279$ (compared to 1.3310 above).  
However, their {\bf ratio}, the average mass per baryon, changes from 
the zeroth approximation by only 0.1\% to $m_{\rm B} = 937.12$~MeV.  
It should be noted that this small difference is entirely driven 
by the slightly different \4he abundance (0.2482 vs. 0.250) and is 
completely consistent with eqs.~5 \& 6; in the ratio of $\rho_{\rm B}$ 
to $n_{\rm B}$, the contributions from D and \3he cancel to a few parts 
in $10^8$.  Thus, from shortly after the quark-hadron transition ($T 
\approx 100$~MeV) until stars form, reinitiating nucleosynthesis, the 
average mass per baryon has decreased by 2.05 MeV, from 939.17 MeV to 
937.12 MeV.

\subsection{Post-Dark Ages}

During the very recent evolution of the Universe, initially small
fluctuations in the mean matter density have been amplified, resulting 
in nonlinear perturbations which collapse under their own gravity, 
separating such regions from the average expansion of the Universe 
and leading to the formation of stars, galaxies, and cluster of 
galaxies.  Prior to the gas being cycled through stars where hydrogen 
is burned to helium and beyond, the mean mass per baryon is preserved 
at its precollapse/post-BBN value.  However, as stellar nucleosynthesis 
proceeds, baryons are incorporated into ever more tightly bound nuclei 
whose binding energy is radiated away.  As a result, locally, the average 
mass per 
baryon decreases, and $m_{\rm B}$ is now distributed inhomogeneously 
throughout the Universe.  So, while the spatially averaged value of
\mb~will be smaller than the post-BBN value, local values of \mb~will
depend on local histories of stellar nucleosynthesis.   As an
illustration we may adopt the recently revised solar abundances
from Asplund, Grevesse and Sauval (2005)~\cite{ags} (AGS) ($X_{\odot} 
= 0.7392$, $Y_{\odot} = 0.2486$, $Z_{\odot} = 0.0122$).  The ratio 
of baryons to hydrogen by mass is $(M_{\rm B}/M_{\rm H})_{\odot} = 
X_{\odot}^{-1}$.  For these values, $y_{4\odot} = 0.08468~(4y_{4\odot} 
= 0.3387$).  Adopting the Geiss and Gloeckler (1998)~\cite{gg} proto-solar 
D and \3he abundances, $(1 + 2y_{2} + 3y_{3} + 4y_{4})_{\odot} = 1.3388$.  
Accounting for the baryons incorporated in all the heavy elements increases 
the ratio of baryons to hydrogen by number to $(N_{\rm B}/N_{\rm H})_{\odot} 
= 1.3554$, leading to $(m_{\rm B}/m_{\rm H})_{\odot} = 0.9981$ and a local 
value of $m_{\rm B\odot}$ = 936.99 MeV.  A word of caution is called for 
here.  The AGS abundances \cite{ags} are photospheric and certainly {\bf not} 
representative of the proto-solar abundances.  A better choice may be the 
BP04+ presolar abundances adopted by Bahcall, Serenelli and Pinsonneault 
(2004)~\cite{bsp} (BSP) ($X_{\odot} = 0.71564$, $Y_{\odot} = 0.26960$, 
$Z_{\odot} = 0.01476$).  In this case, $y_{4\odot} = 0.09486~(4y_{4\odot} 
= 0.37943$); since $(Z/X)_{\odot}$(BSP) = 1.25$(Z/X)_{\odot}$(AGS), the 
heavy element abundances (by number with respect to hydrogen) are scaled 
here by this factor.  For BP04+, $(N_{\rm B}/N_{\rm H})_{\odot} = 
1.40021$, leading to $(m_{\rm B}/m_{\rm H})_{\odot} = 0.99796$ and 
$m_{\rm B\odot} = 936.87$ MeV.  Thus, even in the chemically enriched 
solar vicinity, the average mass per baryon has been reduced from its 
value during the dark ages by only $\sim 0.26$ MeV.

For a BP04+ mix of abundances, along with the assumption that the 
helium abundance increases in proportion to the increase in metallicity
($\Delta Y \propto \Delta Z = Z$), it follows that the average mass per 
baryon {\it decreases} with increasing metallicity as $\Delta m_{\rm B} 
\approx 0.26(Z/Z_{\odot})$ MeV.  This decrease in the average mass per 
baryon mirrors the energy released when hydrogen is burned to helium 
and beyond.  With the simplifying assumption that stellar nucleosynthesis 
occurs at redshift $z_*$, and where $Z$ is the large-scale average 
metallicity, the corresponding present day ratio of the density 
of the energy released to the baryon mass density is
\be
{\Delta\Omega_{\rm B} \over \Omega_{\rm B}} = {\Delta m_{\rm B} 
\over m_{\rm B}}{1 \over (1+z_{*})} \approx {2.7\times 10^{-4}(Z/Z_{\odot}) 
\over (1+z_{*})}.
\ee
For the Spergel \etal \cite{spergel} value of \obh = 0.02233, $\Delta\Omega_{\rm B}
h^2 \approx 6.1\times 10^{-6}(Z/Z_{\odot})(1+z_{*})^{-1}$; for $h = 0.7$, 
this corresponds to $\Delta\Omega_{\rm B} \approx 1.2\times 
10^{-5}(Z/Z_{\odot})(1+z_{*})^{-1}$.  It is interesting to compare 
this to an estimate of the observed energy density in background 
light.  According to Fukugita and Peebles (2004)~\cite{fp}, the sum 
of the optical and far-infrared (FIR) backgrounds is $\Omega_{OPT/FIR} 
\approx 2.4\times 10^{-6}$, so that (see Hauser and Dwek (2001)~\cite{dh})
\be
{\Delta\Omega_{\rm B} \over \Omega_{OPT/FIR}} \approx {5(Z/Z_{\odot}) \over 
(1+z_{*})},
\ee
suggesting that stellar nucleosynthesis should occur at redshifts
$1+z_{*}~\la 5(Z/Z_{\odot})$ or, that $Z/Z_{\odot}~\ga 0.2(1+z_{*})$, if 
the observed background is taken as a lower limit to $\Delta\Omega_{\rm B}$. 

\section{The Baryon Number and Baryon Mass Density Parameters}
\label{eta}

As the Universe expands in the post-BBN era, the number of baryons in a 
comoving volume is conserved and the average mass per baryon is frozen 
at its post-BBN value until stellar nucleosynthesis begins.  Since the 
baryon number and mass densities decrease as the Universe expands, it 
is convenient to express the baryon number and mass densities in terms 
of parameters which remain unchanged as the Universe expands.  The baryon 
{\bf mass} density, $\rho_{\rm B}$, is usually written as a fraction of the 
{\it critical} mass density, defined by $\rho_{c} \equiv 3H_{0}^{2}/8\pi 
G_{N}$, $\Omega_{\rm B} \equiv \rho_{\rm B}/\rho_{c}$.  $H_{0}$ is the 
present value of the Hubble expansion rate parameter, expressed as 
$H_{0} = 100h$~kms$^{-1}$Mpc$^{-1}$.  According to the Particle Data Group 
(PDG)~\cite{pdg}, $\rho_{c} = 1.0537 \times 10^{-5}h^{2}$~GeVcm$^{-3}$.  
The fractional error in $\rho_{c}/h^{2}$, $1.5 \times 10^{-4}$, is 
entirely due to the uncertainty in the experimental value of Newton's 
gravitational constant ($G_{\rm N} = 6.6742 \pm 0.0010\times 
10^{-11}$~m$^{3}$kg$^{-1}$s$^{-2}$~\cite{pdg}).  This uncertainty, while 
not entirely negligible, is subdominant and will be ignored here.  
The ratio of the present baryon mass density, $\rho_{\rm B}$, to 
the hydrogen mass ($m_{\rm H}= 0.938783$~GeV), is related to \obh by
\be
\rho_{\rm B}/m_{\rm H} = 1.1224 \times 10^{-5}~\Omega_{\rm B}h^{2}~{\rm cm}^{-3}.
\ee

It is conventional to compare the baryon {\bf number} density, $n_{\rm B}$, 
to the number density of the Cosmic Background Radiation (CBR) photons, 
$n_{\gamma}$.  To high accuracy this ratio, $\eta_{10} \equiv 
10^{10}(n_{\rm B}/n_{\gamma})_{0}$, is unchanged from the end of BBN 
to the present.  Adopting the present CBR temperature from Mather \etal 
(1999)~\cite{mather}, T$_{\gamma 0} = 2.725 \pm 0.001$K $= 2.725(1 \pm 0.0004)$K,
the corresponding present epoch number density of CBR photons is
\be
n_{\gamma 0} = 410.50({\rm T}_{\gamma 0}/2.725{\rm K})^{3}~{\rm cm}^{-3} 
= 410.50(1 \pm 0.0011)~{\rm cm}^{-3} = 410.50 \pm 0.45~{\rm cm}^{-3}.
\ee
Note that for the same adopted temperature the value listed in~\cite{pdg} 
(410.4), is wrong and has been corrected~\cite{pdg2}.  Combining the 
results above, $\eta_{10}$ may be related to $\Omega_{\rm B}h^{2}$,  
\be
\eta_{10}/\Omega_{\rm B}h^{2} = 273.42(m_{\rm B}/m_{\rm H})^{-1}
(2.725{\rm K}/{\rm T}_{\gamma 0})^{3} = 273.42(1 - 
0.007119{\rm Y_{\rm P}})^{-1}(2.725{\rm K}/{\rm T}_{\gamma 0})^{3}.
\ee
This result is consistent with eq.~4.19 of Serpico \etal (2004)~\cite{serpico};
with the updated value of $G_{\rm N}$~\cite{pdg} their coefficient 273.49 
becomes 273.42.  For T$_{\gamma 0} = 2.725 \pm 0.001$K,
\be
\eta_{10}/\Omega_{\rm B}h^{2} = 273.9 \pm 0.3 + 1.95({\rm Y_{\rm P}} - 0.25).
\ee
This numerical conversion factor (for \Yp = 0.25), $\eta_{10}/\Omega_{\rm 
B}h^{2} = 273.9\pm 0.3$, is a key result of this Letter.  Note that 
even for the solar value of the average mass per baryon derived above 
using the BSP~\cite{bsp} abundances (BP04+), this factor increases only 
very slightly, from 273.9 to a local value of 274.0.  

The ratio of $\eta_{10}$ to \obh can be calculated to an accuracy of 
0.1\%, considerably better than the current 3 -- 4\% accuracy of the 
\obh determination from WMAP.  However, to avoid contributing an 
avoidable error to $\eta_{10}$ as inferred from non-BBN data, this 
coefficient, 273.9, should be employed when converting from \obh to 
$\eta_{10}$.  For comparison, while Trotta and Hansen (2004)~\cite{th} 
use CAMB/RECFast for their numerical results, in their analytic 
results they employ a conversion coefficient closer to 275, which 
differs from the result here by 0.4\%.  The Ichikawa and Takahashi 
(2006)~\cite{it} conversion coefficient of 273.49 (based on Serpico 
\etal\cite{serpico}, using an older value of $G_{\rm N}$) is closer
to the correct result, but still differs from it by more than the 
uncertainty of 0.3.  For the WMAP-Only, 3 year data, Spergel \etal 
(2006)~\cite{spergel} quote (in Table 5) \obh $= 0.02233^{+0.00072}_{-0.00091}$, 
corresponding to $\eta_{10} = 6.116^{+0.197}_{-0.249} \approx 
6.12^{+0.20}_{-0.25}$ ({\bf not} to the value, $\eta_{10} = 6.0965 \pm 
0.2055$, listed in Table 4 of reference~\cite{spergel} which corresponds 
instead to $\eta_{10}/\Omega_{\rm B}h^{2} = 273.0$).  It is this value, 
$\eta_{10} = 6.12$, which was used in \S\ref{postbbn} to calculate the 
primordial abundances needed for a more accurate computation of the 
post-BBN average mass per baryon $m_{\rm B}$.

Finally, it should be noted that the confrontation of cosmological 
models with the CMB data depends on {\it both} the baryon {\it number} 
density (through the number density of electrons) and the baryon 
{\it mass} density.  As a result, such analyses depend on the numerical 
value of the coefficient relating the two.  The codes most commonly 
employed to analyze the CMB data, CAMB~\cite{lcl} and CMBFast~\cite{sz}, 
use RECFAST~\cite{scott}.  The latter assumes that $m_{\rm He}=4m_{\rm H}$, 
so that $m_{\rm B} \equiv m_{\rm H}$ and, it adopts a numerical value 
for $m_{\rm H}$ which is too large by 0.108 MeV.  In addition, the 
codes employ a value for Newton's constant, $G_{\rm N} = 6.67259
\times 10^{-11}$~m$^{3}$kg$^{-1}$s$^{-2}$, which differs slightly 
from the currently recommended best value~\cite{pdg,pdg2}.  As a 
result, the conversion factor built into these codes corresponds to 
$\eta_{10}/\Omega_{\rm B}h^{2} = 273.46$ (provided that T$_{\gamma 0} 
= 2.725$K is adopted), which differs from the more precise result, 
273.91, by 1.5 times the error contributed by the uncertainty in 
T$_{\gamma 0}$.  It should be noted that since Spergel \etal\cite{spergel} 
use CAMB~\cite{lcl} (and RECFAST~\cite{scott}) in their analysis, in 
principle their derived value of $\Omega_{\rm B}h^{2} = 0.02233$ 
corresponds to $\eta_{10} = 6.106 \approx 6.11$.  However, this 
small difference in $\eta_{10}$ doesn't change the BBN-predicted 
primordial \4he abundance (to four significant figures) and, as 
a result, has no effect on the conversion factor calculated here.

\acknowledgments 

Special thanks are due Jim Felten for his careful reading of this 
manuscript and for his constructive suggestions.  I gladly acknowledge 
informative discussions with Richard Cyburt, Don Groom, Kazuhide 
Ichikawa, Antony Lewis, Douglas Scott, David Spergel and Roberto 
Trotta.  This research is supported at The Ohio State University 
by a grant (DE-FG02-91ER40690) from the US Department of Energy.


\end{document}